\documentclass{ws-jai}

\usepackage[latin9]{inputenc}
\usepackage{graphicx}

\usepackage[breaklinks,colorlinks,urlcolor=blue,citecolor=blue,linkcolor=blue]{hyperref}

\usepackage{subfigure}
\usepackage{color}

\begin{document}

%\title[HIPSR Digital Signal Processor]{HIPSR: A Digital Signal Processor for the Parkes 21-cm Multibeam Receiver}

%\author[Price et al.]{D.~C.~Price$^{1,2,3}$\thanks{dprice@cfa.harvard.edu},  A.~Jameson$^{4}$, E.~Carretti$^5$,  W.~van~Straten$^{4}$ L.~Staveley-Smith$^2$,  S.~Schediwy$^1$, M.~E.~Jones$^1$, M.~Bailes$^{4}$, \and et. al. \\
%\affil{$^1$Oxford Astrophysics, Denys Wilkinson Building, Keble Road, Oxford OX1 3RH, United Kingdom}%
%\affil{$^2$International Centre for Radio Astronomy Research, University of Western Australia, Crawley, WA 6009, Australia}%
%\affil{$^3$Harvard-Smithsonian Center for Astrophysics, MS 42, 60 Garden Street, Cambridge, MA 02143, USA}
%\affil{$^4$Swinburne University of Technology, Centre for Astrophysics and Supercomputing Mail H39, PO Box 218, VIC 3122, Australia}
%\affil{$^5$CSIRO Astronomy and Space Science, PO Box 76, Epping, NSW, 1710, Australia}}%
%%
%\jid{PASA}
%\doi{10.1017/pas.\the\year.xxx}
%\jyear{\the\year}

\catchline{}{}{}{}{} % Publisher's Area please ignore

\markboth{D. C. Price}{HIPSR Digital Signal Processor}

\title{HIPSR: A Digital Signal Processor for the Parkes 21-cm Multibeam Receiver}

\author{D.~C.~Price$^{1,2,3,\dagger}$, L.~Staveley-Smith$^{2, 4}$, M.~Bailes$^{4,5}$, E.~Carretti$^{6,7}$,  A.~Jameson$^{4,5}$, M.~E.~Jones$^1$, 
		W.~van~Straten$^{5,8}$, \and S.~W.~Schediwy$^{1,2}$ }

\address{
\small
$^1$Oxford Astrophysics, Denys Wilkinson Building, Keble Road, Oxford OX1 3RH, United Kingdom\\
$^2$International Centre for Radio Astronomy Research, University of Western Australia, Crawley, WA 6009, Australia \\
$^3$Present address: University of California Berkeley, Campbell Hall 339, Berkeley CA 94704\\
$^4$ARC Centre of Excellence for All-sky Astrophysics (CAASTRO) \\
$^5$Swinburne University of Technology, Centre for Astrophysics and Supercomputing Mail H39, PO Box 218, VIC 3122, Australia\\
$^6$CSIRO Astronomy and Space Science, PO Box 76, Epping, NSW, 1710, Australia \\
$^7$Present address: INAF Osservatorio Astronomico di Cagliari, Via della Scienza 5, 09047 Selargius (CA), Italy\\
$^8$Institute for Radio Astronomy \& Space Research, Auckland University of Technology,
Private Bag 92006, Auckland 1142, New Zealand\\
}

\maketitle
\corres{$^\dagger$Corresponding author. Email: \url{dancpr@berkeley.edu}}

% UNCOMMENT THE LINES BELOW IF YOU WISH TO USE BIBTEX
%Citations may be made using the natbib commands \citet{},\citep{} etc.
%%\usepackage[authoryear]{natbib}
%\bibpunct{(}{)}{;}{a}{}{,}
%\setlength{\bibsep}{0.3mm}

%
\begin{abstract}
HIPSR (HI-Pulsar) is a digital signal processing system for the Parkes 21-cm Multibeam Receiver that provides larger instantaneous bandwidth, increased dynamic range, and more signal processing power than the previous systems in use at Parkes. The additional computational capacity enables finer spectral resolution in wideband HI observations and real-time detection of Fast Radio Bursts during pulsar surveys. HIPSR uses a heterogeneous architecture, consisting of FPGA-based signal processing boards connected via high-speed Ethernet to high performance compute nodes.  Low-level signal processing is conducted on the FPGA-based boards, and more complex signal processing routines are conducted on the GPU-based compute nodes. The development of HIPSR  was driven by two main science goals: to provide large bandwidth, high-resolution spectra suitable for 21-cm stacking and intensity mapping experiments; and to upgrade the Berkeley-Parkes-Swinburne Recorder (BPSR), the signal processing system used for the High Time Resolution Universe (HTRU) Survey and the Survey for Pulsars and Extragalactic Radio Bursts (SUPERB).
\end{abstract}

\section{Introduction}

The CSIRO Parkes 64-m radio telescope is equipped with a multibeam receiver that covers a frequency range of 1.23-1.53 GHz. This receiver consists of 13 cryogenically cooled dual-polarization feeds in a hexagonal array \citep{staveley1996}. The receiver was designed primarily for redshifted 21-cm hydrogen line (HI) observations over 0$\le$z$<$0.2 and was installed in 1997, along with the MBCORR digital signal processing (DSP) system. Until its decommissioning in 2016, MBCORR remained the primary signal processing system for 21-cm line emission studies with the multibeam receiver.

An additional DSP backend was installed at Parkes in 2008, designed for high time-resolution spectroscopy with the multibeam receiver \citep{mcmahon2008, hitrun1}. This backend was known as the Berkeley-Parkes-Swinburne Recorder (BPSR) and it replaced an analog filterbank that was designed and used for the Parkes Multibeam Pulsar Survey \citep[PMPS;][]{slc+99v2}. BPSR originally consisted of 26 Xilinx Virtex-2 Pro FPGA-based spectrometers connected by 10 gigabit Ethernet to a cluster of 13 servers, which further processed and reduced the signal before writing the spectra to disk for offline processing. BPSR processes the entire usable bandwidth of the receiver at high time resolution and has enabled a remarkable series of discoveries of radio pulsars and fast radio bursts \citep[e.g.][]{kjv+10,lbb+10,bbb+11,bbb+11a,bbj+11,kjb+12,bjb+12,bbb+12,tsb+13,bbb+13,lbb+13,tjb+13,nbb+14,btb+15}.  However, the spectra produced by BPSR are too coarsely channelized ($\Delta\nu\sim$390 kHz) for neutral hydrogen observations.

In this article, we report on the HI-Pulsar (HIPSR) system, an upgrade to BPSR providing new high spectral resolution modes for HI observations. The HIPSR system, installed at Parkes in 2011, digitizes the entire usable bandwidth of the multibeam receiver, has increased resilience to radio frequency interference, and provides a significant increase in overall computational power in comparison to MBCORR and BPSR. 

\section{System Design}

The HIPSR system uses an heterogeneous architecture  consisting of an interconnected network of general purpose graphics processing units (GP-GPUs) and field programmable gate arrays (FPGAs). This architecture allows processing tasks to be divided between DSP `nodes', with wide-bandwidth DSP conducted on FPGAs, and moderate to low-bandwidth DSP tasks performed on the GPUs. 

This division of DSP is motivated by the relative advantages and weaknesses of the two DSP platforms. In recent years, GPUs have emerged as
an attractive platform for parallelizable radio astronomy algorithms. For example, pulsar dedispersion pipelines \citep{barsdell2010, magro2011}, and low to moderate-bandwidth, large-N correlators \citep{clark2013, kocz2014}, have been shown to be well suited to GPU implementations. 
% Nevertheless, the input/output (I/O) bandwidth of GPUs is a bottleneck that limits their performance for many wide-bandwidth radio astronomy applications.  [WvS disagrees; I/O to a GPU is limited only by the PCI bus and the ability of the programmer]

For wide-bandwidth DSP, FPGAs are ubiquitous. FPGAs are well suited to wide-bandwidth applications as their I/O bandwidth is limited only
by the number and speed of I/O pins upon the chip, whereas the I/O bandwidth of CPUs and GPUs are limited by the speed of the host's data busses. On the other hand, algorithms which require large amounts of data to be stored in memory are poorly suited to FPGA-based implementations, as there is a limited amount of logic and memory available within each FPGA chip.

Many of the algorithms used in radio astronomy are well suited to FPGA implementations. For example, the CASPER collaboration provides a library of efficient FPGA gateware for radio astronomy, such as digital downconverters, vector accumulators, and polyphase filterbank (PFB) structures \citep{parsons2009,hickish2016}. Pulsar dedispersion pipelines have also been demonstrated on FPGAs \citep{clark2014}. The decision between GPUs and FPGAs (or any other DSP architecture for that matter) is dependent mainly upon the ease of deployment for the set of algorithms one wishes to run. For multi-purpose DSP systems --- such as HIPSR --- a heterogeneous architecture allows the advantages of both platforms to be leveraged.

\begin{figure}
\begin{centering}
\includegraphics[width=0.7\columnwidth]{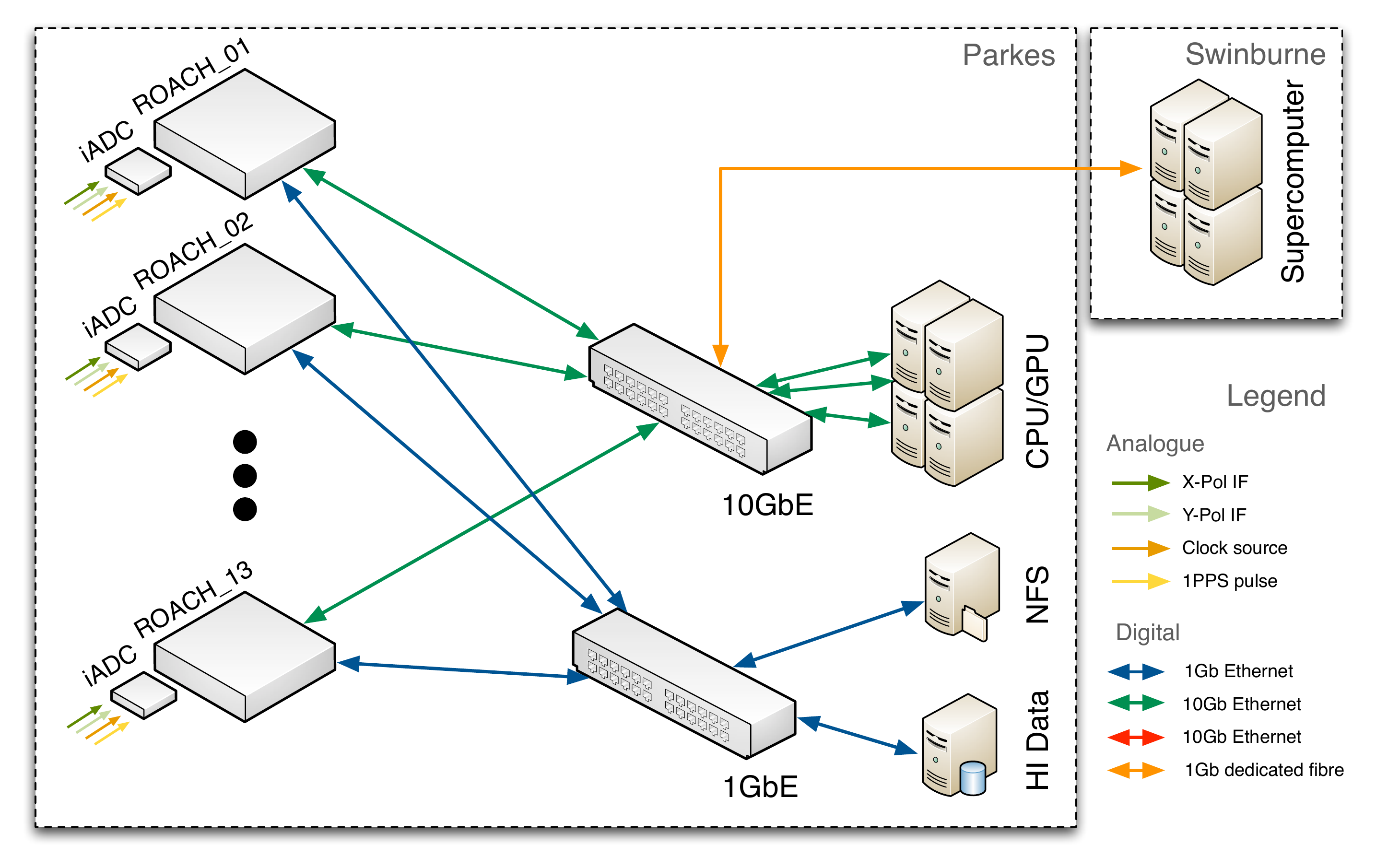}
\par\end{centering}

\caption{HIPSR network diagram. The ROACH-based DSP frontend is pictured on
the left; these are connected via a 10~Gb Ethernet switch to the GPU-based DSP
backend. 1~Gb Ethernet is used for monitor and control, and for low-bandwidth data transport from the FPGA frontend.\label{fig:HIPSR-network-diagram}}
\end{figure}

\subsection{HIPSR Packetized Architecture}

The HIPSR system may be described as a multi-node DSP `frontend' connected to a multi-node DSP `backend' by a 10Gb Ethernet (10GbE) switch. 
The FPGA-based DSP frontend conducts signal processing tasks such as signal filtering and channelization; the GPU-based DSP backend
performs higher-level, moderate- to low-bandwidth DSP such as running pulsar processing pipelines. A diagram of the system architecture is
shown in Fig.~\ref{fig:HIPSR-network-diagram}.

Packetized DSP systems are becoming increasingly common within radio astronomy, see for example the Precision Array for Probing the Epoch of Reionization (PAPER) correlator \citep{parsons2009b}, the Large-Aperture Experiment for detecting the Dark Ages (LEDA) correlator \citep{kocz2014b}, and VEGAS, the Versatile Greenbank Astronomical Spectrometer \citep{roshi2011, chenna2014}. The inclusion of a switch greatly increases the flexibility of the system as data can be dynamically routed between nodes. It does, however, require all data to be packetized along with metadata that describes the  packet's data payload. Nevertheless, using a switched 10GbE-based packetized architecture instead of defining and commissioning a bespoke backplane has drastically decreased development time: all interconnect is off-the-shelf, industry standard, commodity hardware.

\subsection{Hardware Overview}

The DSP frontend consists of 13 Reconfigurable Open Architecture Computing Hardware (ROACH) signal processing boards, each of which has an FPGA as its centerpiece. Each ROACH has a 2-input, 1~Gsample/s digitizer mezzanine card installed, which allows for a total of 26 inputs to be digitized (i.e. every IF of the multibeam receiver). The DSP backend consists of 8 server nodes, each of which is equipped with dual Nvidia C2070 GPUs. This hardware is discussed in further detail below.

\subsubsection{The CASPER ROACH Board\label{sub:The-CASPER-ROACH}}

The ROACH is a Xilinx SX95T Virtex-5 FPGA-based signal processing board designed for radio astronomy applications. ROACH is open-source hardware provided by the Collaboration for Astronomical Signal Processing and Electronic Research (CASPER)\footnote{\url{http://casper.berkeley.edu}}. 

The ROACH board has four CX4-type Ethernet connectors that allow up to 40~Gb/s of bandwidth for data transfer over 10~GbE or XAUI protocols. Daughter boards, such as analog-to-digital converters, may be connected via the two Z-DOK interfaces, each with 40 differential pairs connecting to the FPGA I/O pins. General Purpose Input/Output (GPIO) pins are also provided for control of miscellaneous sensors and external devices. For temporary storage of large vectors, the FPGA is connected to quad data rate random access memory (QDRII+ RAM) and double data rate memory (DDR2). 

The Virtex-5 FPGA on the ROACH is connected to a PowerPC chip by an On-Chip Peripheral Bus (OPB) controller. The PowerPC runs a customized Linux kernel and provides monitor and control for the board. It has its own dedicated DDR2 and Flash memory, 1~Gb Ethernet port, USB port, and SD memory card socket. The OPB interface allows the PowerPC to reprogram the FPGA and to access registers and memory resources on the FPGA fabric. This simplifies data readout and provides a convenient way to reconfigure the firmware running on the FPGA.

\subsubsection{iADC Digitizer Card\label{sub:iADC-digitizer-card}}

The digitizer used in HIPSR, known as the iADC or ADC2x1000-8, is an 8-bit, 2-input analog-to-digital converter (ADC) card designed by CASPER. The iADC is a Eurocard-sized daughter board that connects to the ROACH via the Z-DOK connector. It is powered by an Atmel/e2V AT84AD001B digitizer chip, which runs at speeds of up to 1024~Msample/sec. Inputs are connected to the iADC's SMA jacks; a Mini~Circuits ADTL2-18 balun (30 MHz to 1.8 GHz bandwidth) converts the signals from single-ended to differential. 

When in operation, an external frequency reference tone must be presented as a clock signal. A pulse per second (1PPS) may also be connected; this signal is passed through to the FPGA so multiple ROACH boards may be synchronized to within a FPGA clock cycle. Using the iADC, two signals with bandwidths of up to 512 MHz may be digitized. Alternatively, the AT84AD001B offers an interleave mode which sacrifices one input but is then able to digitize 1024 MHz from a single input. In HIPSR, the former is used exclusively.

Two important characteristics for digitizers are crosstalk levels between signal inputs, and the linearity of the digitizer response. Crosstalk of the iADC has been measured by to be under $-28$~dB \citep{parsons2009b}, and was found to stable enough such that it can be removed to levels $\sim60$ dB. Differential and integral non-linearity has been tested experimentally and found to be in accordance with the manufacturer's quoted specification (differential 0.25 LSB, integral 0.5 LSB). Further specifics of the iADC are listed in Tab.~\ref{tab:Atmel-AT84AD001B-performance}.

\begin{table}
\caption{\label{tab:Atmel-AT84AD001B-performance}Atmel AT84AD001B performance
characteristics \citep*{atmel2009}.}

\centering{}%
\begin{tabular}{lr}
\hline 
{\small ADC specification} & \tabularnewline
\hline 
\hline 
{\small Sampling rate} & {\small 1GS/s }\tabularnewline
 & {\footnotesize (2GS/s interleaved)}\tabularnewline
{\small Number of bits} & {\small 8}\tabularnewline
{\small Effective number of bits} & {\small 6.8 }\tabularnewline
{\small Full scale input } & {\small 500mVpp}\tabularnewline
{\small Spurious free dynamic range} & {\small $-54$ dBc}\tabularnewline
{\small Differential non-linearity} & {\small 0.25 LSB}\tabularnewline
{\small Integral non-linearity} & {\small 0.5 LSB}\tabularnewline
{\small Bit error rate} & {\small $\mbox{10}^{-13}$ at 1GS/s}\tabularnewline
\hline 
\end{tabular}
\end{table}

\subsubsection{GP-GPU Server Nodes}

The GP-GPU server nodes are bespoke systems, built to specification by Silicon Graphics Pty. Ltd. Each of the 8 server nodes comprise dual 2.66 GHz Intel Xeon six-core CPUs, 48 GB DDR3 memory, and dual Nvidia Tesla C2070 GP-GPUs. While a majority of DSP is conducted on the GPUs, the Intel Xeon CPUs may also be used, if required. 

As each server has a single CX4-type 10~GbE network interface card (NIC), the input data rate for each server is limited to 10~Gb/s. Given that there are 26~IFs which must be processed by 8 server nodes, the DSP frontend must decrease the data rate \footnote{digitized data rate is $2\times400\mbox{MHz}\times8\mbox{bit}=6.4\mbox{Gb/s}$} from each IF from 6.4~Gb/s to about 3.0~Gb/s.

\subsubsection{Data Server}

In addition to the GPU server nodes, there is a server that provides monitor and control for the DSP frontend, and provides data storage. This server has a 2.4 GHz Intel Xeon four-core CPU, and 24 GB DDR3 memory. As it is not used for DSP, it does not have a GPU. This server runs a DHCP daemon and provides a network file system (NFS) to the ROACH boards.

\subsubsection{Node Interconnect}

As shown in Fig.\ \ref{fig:HIPSR-network-diagram}, the DSP frontend and backend nodes are connected via a 10~Gb Ethernet switch. This Cisco 4900M switch is configured with 24 CX4-type ports, 13 of which are connected to the DSP frontend ROACH boards, and 8 of which are connected to the server nodes. The switch allows bi-directional data flow between nodes. The 10~GbE switch also has a dedicated 1~GbE fiber link to the Green II supercomputer \footnote{\url{http://astronomy.swin.edu.au/supercomputing/green2/}} at Swinburne University of Technology in Melbourne, Australia.

All nodes are also connected by 1~GbE through a Cisco 3750 switch. For low data bandwidth applications, where all of the necessary DSP to reduce data rates is conducted on the FPGA, all data may be read off the 1~GbE ports of the ROACH boards.

\subsection{Installation and Cabling}

\begin{figure}
\begin{centering}
\includegraphics[width=0.6\columnwidth]{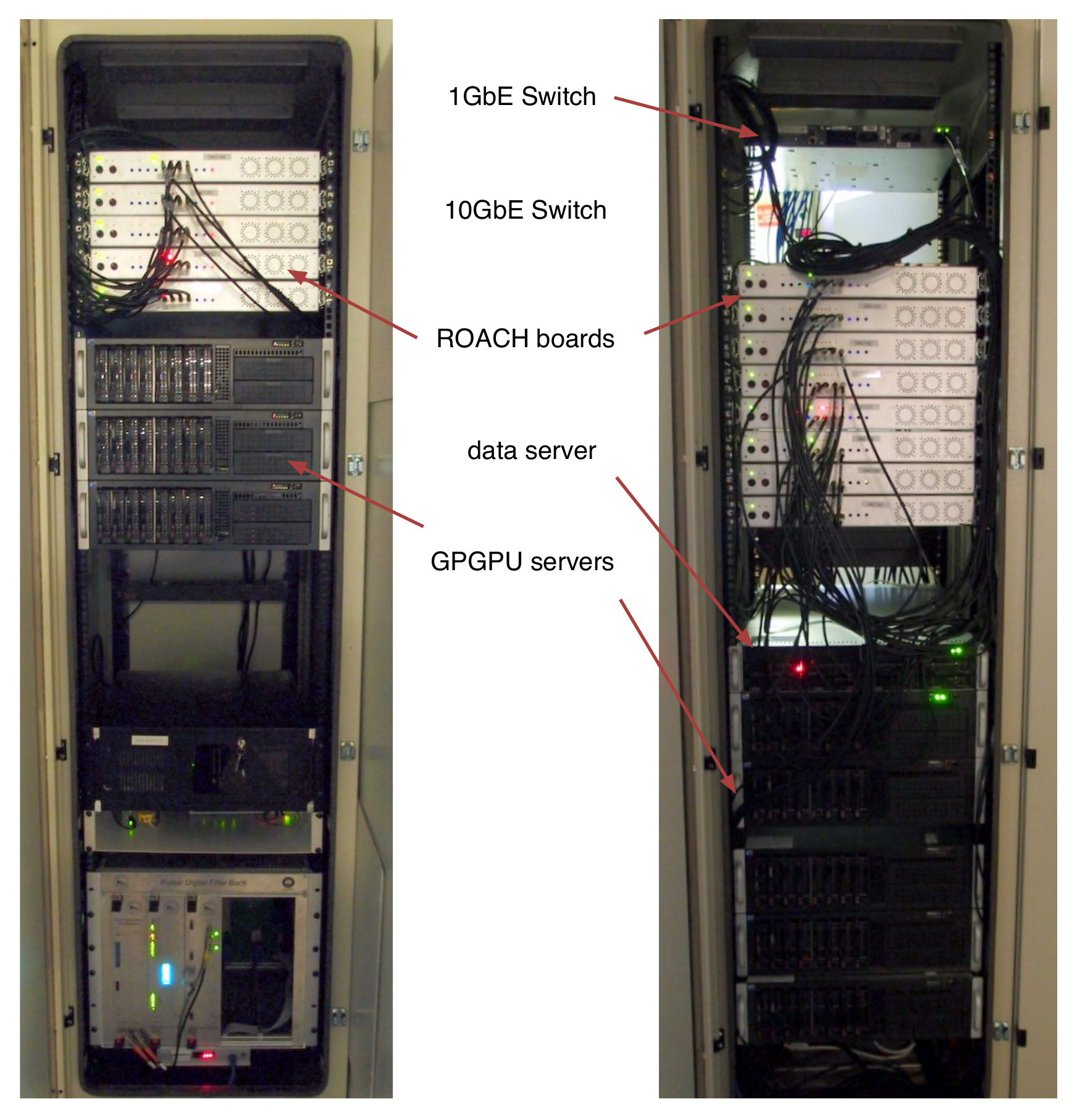}
\par\end{centering}

\caption{HIPSR hardware installed into Parkes 64-m telescope RFI shielded racks.
\label{fig:HIPSR-racks}}
\end{figure}

In March 2012, the HIPSR hardware was installed into radio frequency interference (RFI) shielded racks (Fig.~\ref{fig:HIPSR-racks}), located on the second floor of the Parkes 64-m telescope building (this room lies underneath the telescope dish). A set of BNC-terminated cables were laid under the floor from the existing multibeam IF distribution panel to feedthroughs installed on the racks; another internal set of cables connect the bulkhead to the SMA input of the iADC cards. The 1PPS input of the iADC cards are connect to a PPS distribution unit, and similarly a synthesized reference clock signal is distributed to each iADC board. A single `master' ROACH board is connected to the Parkes Calibration Control Unit (CCU), by a coaxial cable between the CCU and a GPIO output on the master ROACH.

Each server node and ROACH board is connected to a 1~GbE switch by an Ethernet cable (CAT-5E STP). Similarly, each server node and ROACH is connected to a 10~GbE switch via CX4-type cables.

\section{Modes of Operation}

HIPSR can be operated in either high-time resolution or 21-cm spectral line modes. While commensal modes are possible in principle, spectral line work requires the cycling of a reference noise diode at 128 Hz, which is problematic for high-time resolution studies. As such, commensal observations are not supported.

\subsection{High Time Resolution Modes}

The total data rate as digitized for the 26 inputs to the ROACH iADC cards is 166.4~Gb/s.  As there are a total of 8x 10~GbE connections to the GPU servers, bandwidth over the 10~GbE network is a bottleneck. Therefore, the data rate is typically decreased by time integration on the FPGA boards, lowering bit depth, extracting a subset of frequency channels, or dropping the number of IF signals sent to the GPU servers.

Data are captured on the GPU servers using {\tt PSRDADA}\footnote{\url{http://psrdada.sourceforge.net/}}, which provides the `pipeline' infrastructure on which several modes of operation are based.  That is, {\tt PSRDADA} writes the data to ring buffers in shared memory and manages the routing, monitoring and processing of the high-throughput data stream. The various pulsar observing modes are detailed in \citet{sarkissian2011}; we give a brief overview of these modes here, focusing on the modifications and improvements afforded by HIPSR.

%The High Time Resolution Universe survey (HTRU) is an ongoing pulsar
%and fast transient survey using the Parkes multibeam receiver on the
%64m telescope. It has already discovered a slew of pulsars, dynamic
%radio transients and most recently, provided compelling evidence for
%a planet made out of diamond \citep{Keith2010, BurkeSpolaor2011, Bailes2011} . 

%Signal digitization, channelization and packetization for the HTRU
%was originally performed on first generation CASPER hardware \citep{Mcmahon2008}.
%This functionality has been reproduced on ROACH by porting the FPGA
%firmware written for BPSR to Virtex-5. While some modest improvements
%have been made to the FPGA firmware (more poly-phase filterbank taps
%and improved bit precision), the main benefit to the HTRU survey ---
%and to pulsar science with Parkes --- is from the increased processing
%power provided by a new CPU/GPU beowulf cluster which will replace
%the BPSR servers. 

\subsubsection{Berkeley-Parkes-Swinburne Recorder (BPSR)}

The aforementioned Berkeley-Parkes-Swinburne Recorder (BPSR) is a high time-resolution digital filterbank data acquisition and processing system, designed for pulsar and transient detection. When originally installed, BPSR consisted of 13 CASPER iBOB boards, each connected to  a Dell 1950 server-class machine. The functionality of the original BSPR system is completely reproduced and BPSR is now a mode of operation of the HIPSR system. This required a recompilation of the iBOB FPGA code for the ROACH platform, during which some important improvements to the polyphase filterbank were made; first, the ROACH enabled improved bit-precision within the FFT and the addition of extra filterbank taps; more importantly, the spectrometer was upgraded to produce all four Stokes parameters, which enabled the first detection of the polarization of a Fast Radio Burst \citep[FRB;][]{petroff2015}.
 
For BPSR, the ROACH boards apply a 1024-channel, 4-tap Hamming-windowed polyphase filterbank to the 400~MHz bandwidth inputs. The signals are then detected, integrated for 64~$\mu$s, decimated to 8~bits, then packetized and sent via 10~GbE to the GPU compute nodes. The GPU compute nodes then sum, normalize, and decimate to 2 bits per sample, before writing to disk in real-time using {\tt DSPSR}\footnote{\url{http://dspsr.sourceforge.net/}} \cite{vb11}.

On HIPSR, the BPSR system is capable of realtime searches for FRB events using the {\tt Heimdall}\footnote{\url{http://heimdall-astro.sourceforge.net/}} processing software. {\tt Heimdall} performs a search for single pulses across a range of specified dispersion measures, returning a list of candidate events that may then be saved to disk for further analysis.  The {\tt Heimdall} pipeline employs the {\tt dedisp}\footnote{\url{http:://dedisp.googlecode.com}} library, which implements a GPU-accelerated algorithm for incoherent dedispersion \cite{Barsdell2012}. This real-time ability was exploited in \citet{petroff2015}, where a FRB event was detected and used as a trigger for follow-up observations at different observatories.

The BPSR system is discussed in further detail in \citet{hitrun1} and \citet{mcmahon2008}. Note that these papers detail the original iBOB-based hardware; however, apart from the additional features described above, these papers remain an accurate overview.

\subsubsection{CASPER Parkes Swinburne Recorder}

The CASPER Parkes Swinburne Recorder (CASPSR) is used to perform phase-coherent dispersion removal over the full bandwidth of a single dual-polarization beam. CASPSR consists of a single iBOB and iADC digitizer that digitizes and outputs the full 400 MHz of bandwidth at Nyquist rate (6.4~Gb/s) to the backend servers over 10~GbE. These data are then saved to disk, using {\tt PSRDADA} for packet capture and {\tt DSPSR} for disk write. While the iBOB firmware has been ported to ROACH, the original iBOB frontend is still primarily used.

\subsection{21-cm line spectral observing modes}

\subsubsection{Wide-bandwidth Spectroscopy}

For the 21-cm modes of operation all DSP is performed on the ROACH boards.  {\tt HISPEC\_400\_8192} is an 8192-channel, 400-MHz bandwidth PFB spectrometer, which computes all Stokes parameters. That is, each ROACH board processes the polarization pair from a single beam of the receiver and computes the auto-correlation and cross-correlation products for that pair. To use this spectrometer with the multibeam system, 13 copies of this spectrometer are run in parallel, each on a separate ROACH board. In comparison to the MBCORR autocorrelation spectrometer (ACS\footnote{see \citet{price2016} for a comparison of ACS and PFB spectrometer implementations.}), {\tt HISPEC\_400\_8192} offers a factor of 6 increase in bandwidth while still achieving higher spectral resolution, vastly improved inter-channel isolation, and increased dynamic range from 8-bit sampling. A direct comparison to MBCORR is shown in Tab.~\ref{tab:HISPEC}.

Due to its increased instantaneous bandwidth,  {\tt HISPEC} can yield significantly faster survey speeds (a factor of 2-3 times) than MBCORR for wide-bandwidth 21-cm line surveys (i.e. $>$64 MHz). This will decrease the required observing time for 21-cm intensity mapping experiments as proposed by \citet{li2012}, and for spectral stacking such as that by \citet{delhaize2012}. Supporting such science projects was a key motivation for the HIPSR system.

\subsubsection{Mid-bandwidth Spectroscopy}

A 16384-channel, 200-MHz bandwidth mode ({\tt HISPEC\_200\_16384}) for 21-cm observations was also developed. This mode is similar in functionality to {\tt HISPEC\_400\_8192}, but halves the bandwidth and doubles the number of channels; this trade-off is required as there are limited resources within the FPGA. {\tt HISPEC\_200\_16384} does not compute cross terms between polarizations, so Stokes parameters are not computed.

All HIPSR modes digitize at a Nyquist rate of 800 Msamples/s. In {\tt HISPEC\_200\_16384}, a half-band digital filter and complex mixer perform digital downconversion to select out 200 MHz of bandwidth, after which the 16384-channel PFB is applied.  A comparison of {\tt HISPEC} modes is shown in Tab.~\ref{tab:HISPEC}.

\subsubsection{Noise Adding Radiometer Calibration}

HISPEC employs noise adding radiometer (NAR) techniques to accurately measure system temperature \citep{brunzie1988}. NAR techniques allow for greater calibration accuracy to be achieved, at the expense of increasing system temperature (increasing $T_{sys}$ by 5-10\%). In a NAR, a calibrated noise source is turned on and off at $\sim$100 Hz to inject a known signal on top of the sky signal. Calibration is then achieved by monitoring the power difference between on and off states; this differential measurement is less susceptible to bandpass gain variations than an absolute power measurement. 

In the 21-cm multibeam receiver, a calibrated noise source is injected after the orthomode transducer (part of the antenna feed); this noise source is controlled through the CCU. In HISPEC, a master ROACH is connected to the CCU, and all other ROACHes are synchronized to within one clock cycle (5ns) of the master through the distributed 1PPS. The  HISPEC firmware on each ROACH is then able to compute two values: $P_{\rm{on}}$, the total power of the signal when the noise source is switched on; and $P_{\rm{off}},$ the total power of the signal when the noise source is off. The relationship between system temperature $T_{\rm{sys}}$ and the noise source's temperature $T_{C}$ is then given by  

\begin{equation}
T_{\rm{sys}}=\frac{T_{C}}{\frac{P_{\rm{on}}}{P_{\rm{off}}}-1}.
\end{equation}
The value of $T_{C}$ may be measured as required by comparison to
astronomical calibration sources such as 1934-638 and Hydra A, which
have well characterized flux models.

%\begin{table}
%\caption{Comparison of {\tt HISPEC\_400\_8192} with MBCORR 64MHz mode.\label{tab:MBCORR-HIPSR}}

%\centering{}%
%\begin{tabular}{lcc}
%\hline 
%Specification & HISPEC & MBCORR\tabularnewline
%\hline 
%%\hline 
%Bandwidth (MHz) & 400  & 64 \tabularnewline
%Sampling (levels) & 256  & 3 \tabularnewline
%Number of channels & 8192 & 1024\tabularnewline
%Spectral resolution (kHz) & 48.8 & 62.5\tabularnewline
%Velocity resolution (km/s) & 10.3 & 13.4 \tabularnewline
%Implementation & PFB & ACS\tabularnewline
%Interchannel isolation & $\sim$50dB & $\sim$6.6dB\tabularnewline
%Spectral products & {\footnotesize $\begin{array}{c}
%\mbox{XX}^{*}\mbox{YY}^{*}\end{array}$} & {\footnotesize $\mbox{XX}^{*}$$\mbox{YY}^{*}$}\tabularnewline
% & {\footnotesize $\begin{array}{c}\mbox{XY}^{*}\mbox{YX}^{*}\end{array}$} & -- \tabularnewline%

%Integration length (s) & 2-5 & 2-5\tabularnewline
%\hline 
%\end{tabular}
%\end{table}

\begin{table}
\footnotesize
\caption{Comparison of HIPSR 21-cm line observing modes with MBCORR 64-MHz mode.\label{tab:HISPEC}}

\centering{}%
\begin{tabular}{lccc}

\hline 
& {\tt HISPEC} & {\tt HISPEC}  &  {\tt MBCORR} \tabularnewline
Specification & {\tt 400\_8192} & {\tt 200\_16384} & {\tt 64\_1024}\tabularnewline

\hline 
\hline 
Bandwidth (MHz) & 400  & 200 & 64\tabularnewline
Sampling (bits) & 8  & 8 & 3 \tabularnewline
Number of channels & 8192 & 16384 & 1024 \tabularnewline
Spectral res. (kHz) & 48.8 & 12.2 & 62.5 \tabularnewline
Velocity res. (km/s) &  10.3 &  2.5 & 13.4 \tabularnewline
Implementation & PFB & PFB & ACS\tabularnewline
 -- Number of taps & 4 & 4 & - \tabularnewline
 -- Window function & Hamming & Hamming & -\tabularnewline
Interchannel isolation & $\sim$50~dB & $\sim$50~dB & $\sim$6.6dB\tabularnewline
Spectral products & {\footnotesize $\begin{array}{c}
\mbox{XX}^{*}\mbox{YY}^{*}\end{array}$} & {\footnotesize $\mbox{XX}^{*}\mbox{YY}^{*}$}
& {\footnotesize $\mbox{XX}^{*}\mbox{YY}^{*}$}\tabularnewline
 & $\mbox{XY}^{*}\mbox{YX}^{*}$ & & \tabularnewline
Integration length (s) & 2-5 & 2-5 & 2-5\tabularnewline
\hline 
\end{tabular}
\end{table}

\section{Results}

%To validate and test HISPEC modes, observations of extragalactic HI emission were undertaken and the resultant spectra were compared against measurements taken with MBCORR. To test the upgraded BPSR modes, a pulsar (Vega) with known properties was observed. These initial observations are discussed below.

\subsection{System bandpass}
\begin{figure}
\begin{centering}
\subfigure[Full bandwidth]{
	\includegraphics[width=0.6\columnwidth]{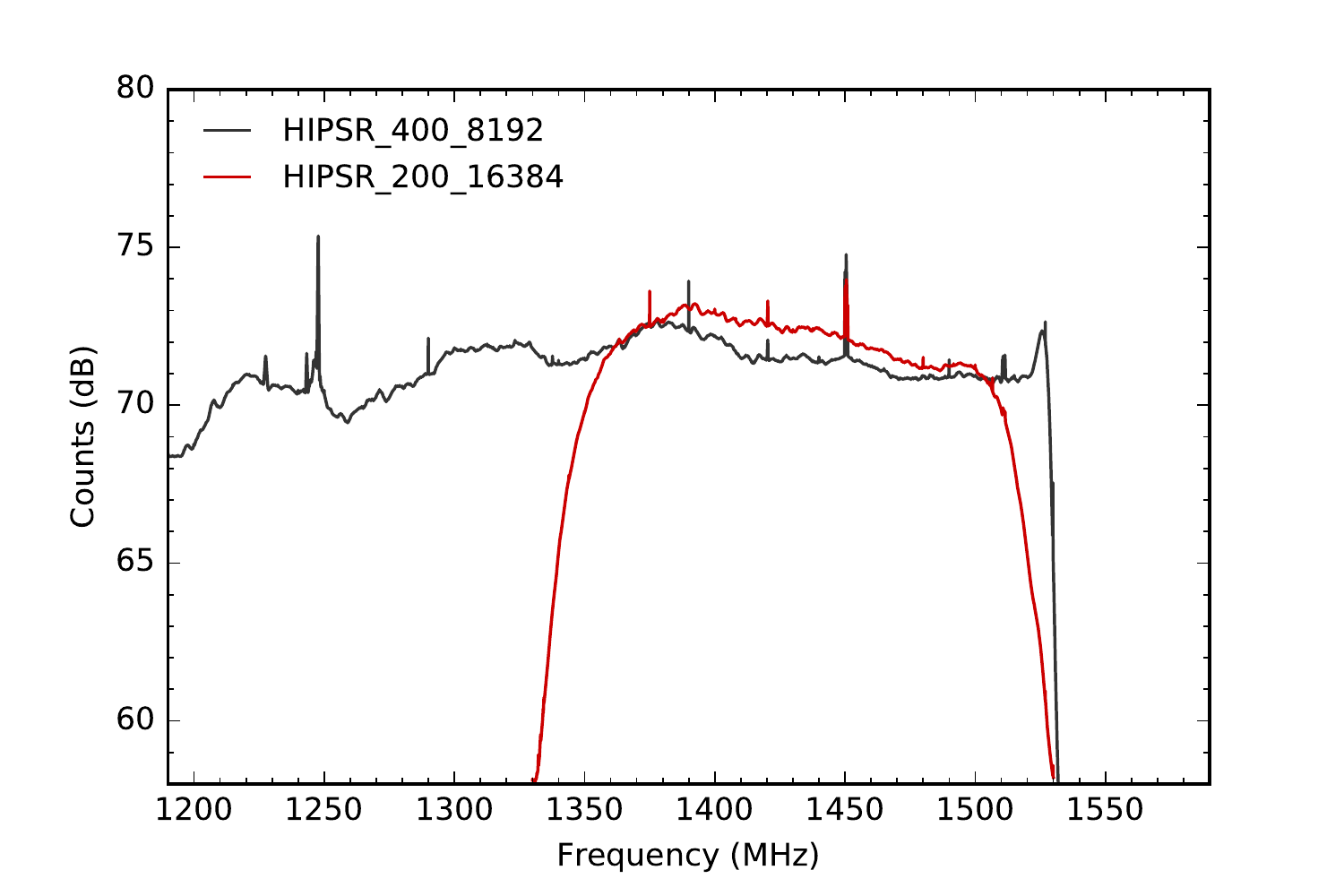}
    \label{fig:HIPSR-comparison-full}
}

\subfigure[Zoom of galactic 21-cm line profile]{
	\includegraphics[width=0.6\columnwidth]{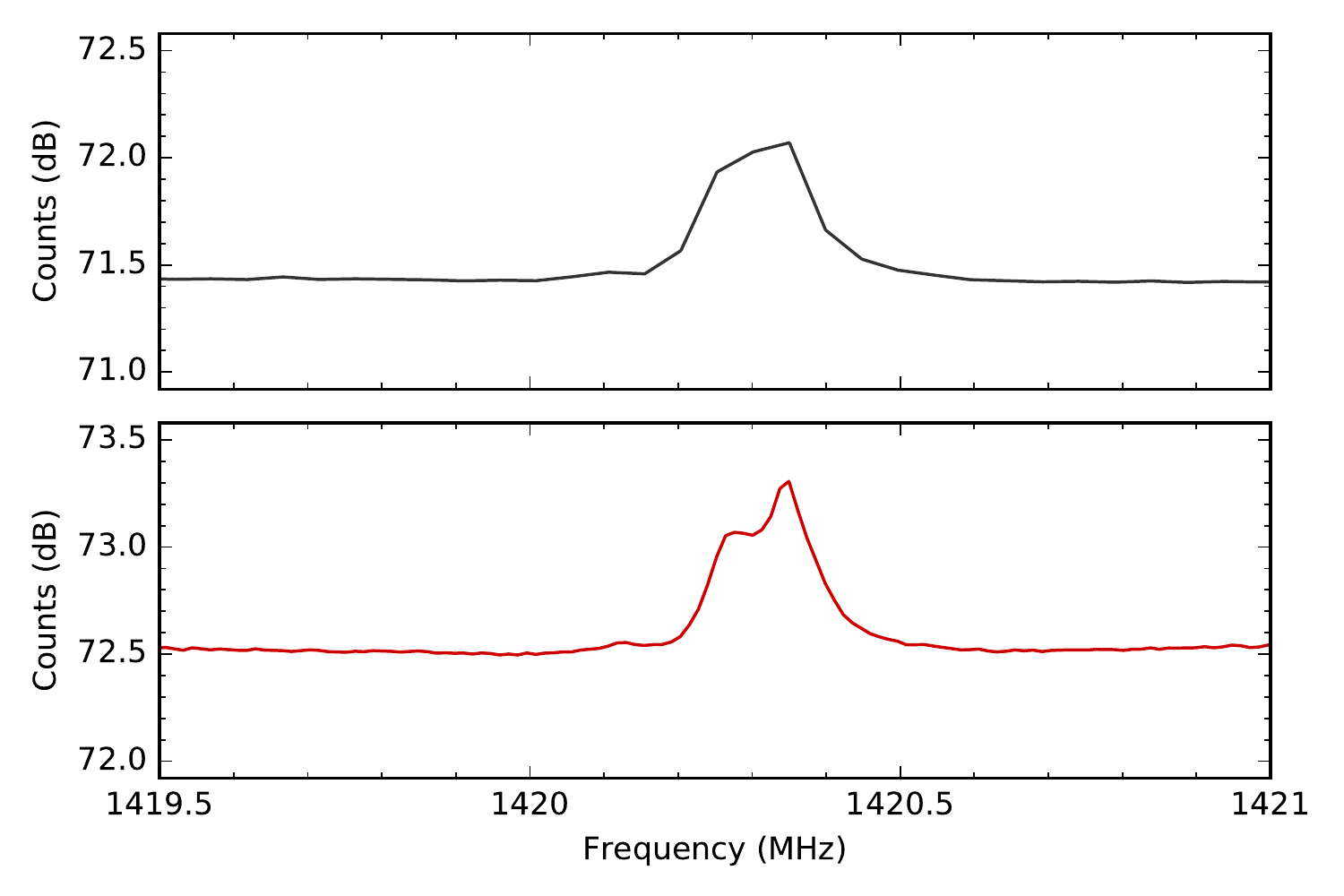}
    \label{fig:HIPSR-comparison-zoom}
}

\caption{Example spectra for 400~MHz (black) and 200~MHz (red) 21-cm operating modes. Data were taken sequentially during system testing on 30/05/2014, with the telescope pointed at the calibrator source 0407-638. 
	\label{fig:HIPSR-comparison}
}

\end{centering}
\end{figure}

Fig.~\ref{fig:HIPSR-comparison} shows example power spectra for an observation of calibrator source 0407-638 on 30 May 2014 using the two HISPEC modes. These tests used the central beam of the receiver and its corresponding analog signal chain. The analog chain performs signal conditioning to amplify, downmix and filter the signals to provide a suitable baseband signal (IF) to the digitizer; the analog chain mixes the receiver signal (RF) with a local oscillator (LO) and a sideband is extracted using analog filters with a 60-400~MHz passband. For these data, the LO was set to extract the lower sideband and as such the IF is reversed in frequency.

Several features are apparent in Fig.~\ref{fig:HIPSR-comparison-full}. The filter response of the {\tt HIPSR\_200\_16384} half-band filter can be seen, and the sharp cutoff of the analog chain's 60~MHz highpass filter can be seen at the upper end of the {\tt HIPSR\_400\_8192} spectrum (IF is frequency reversed). A periodic ripple is apparent in both spectra, caused by standing waves between the dish surface and the focus. Several sources of radio interference are visible as spikes across the band (e.g. 1227~MHz and 1450~MHz). Fig.~\ref{fig:HIPSR-comparison-zoom} shows a zoom-in on the galactic 21-cm line profile for the 0407-638 observation. The {\tt HIPSR\_200\_16384} spectrum resolves a double-peak in the line profile that cannot be seen in the {\tt HIPSR\_400\_8192} profile. 

\subsection{Radio interference}

\begin{figure}
\centering{}\includegraphics[width=0.6\columnwidth]{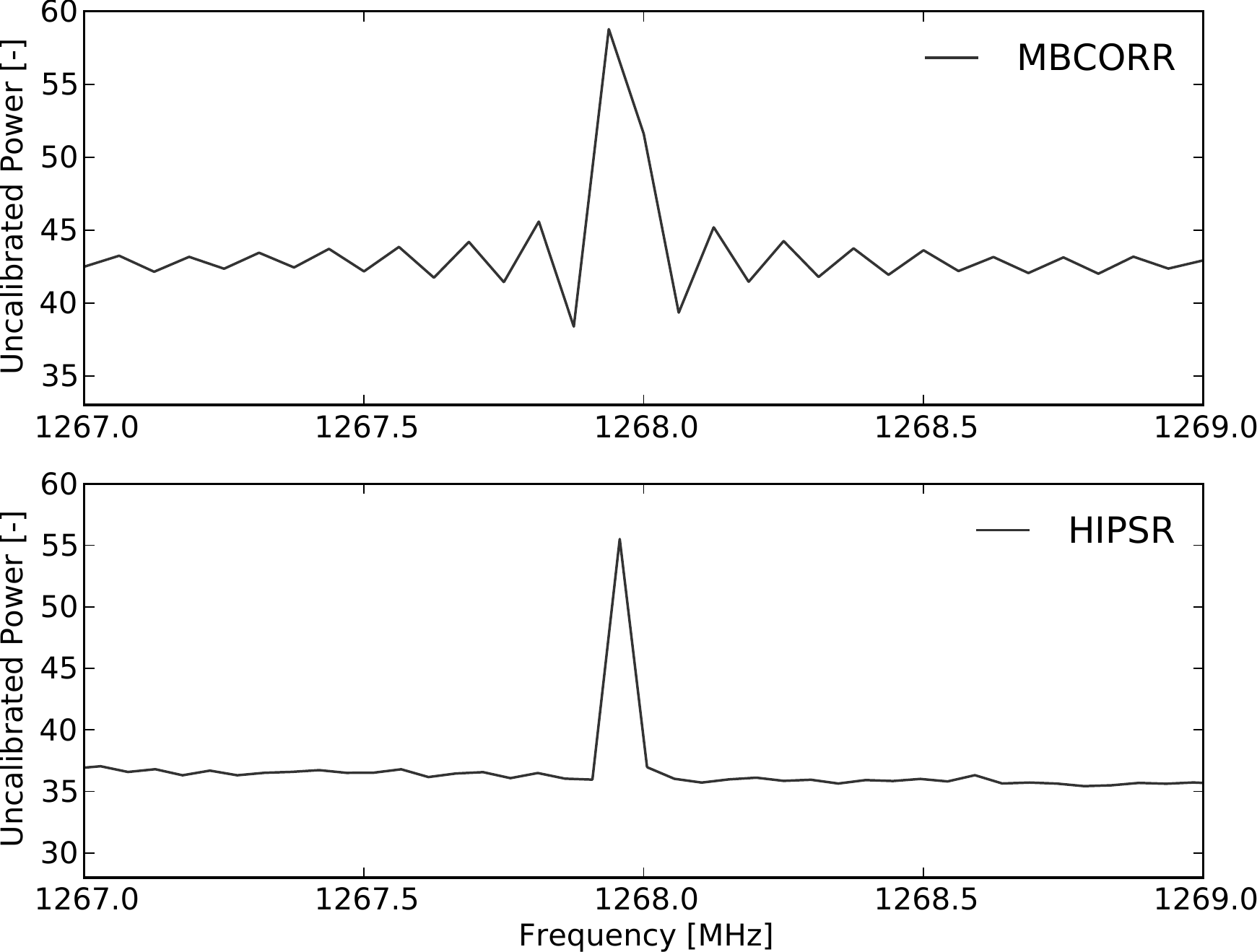}\caption{Example of HIPSR and MBCORR response to narrowband RFI. The (unsmoothed) data output from MBCORR is visibly affected by Gibbs ringing.\label{fig:Galactic-HI}}
\end{figure}

Compared to MBCORR, HIPSR has far superior interchannel isolation ($\sim50$~dB vs. $\sim6.6$~dB between neighbouring channels). As a result, spectra are less affected by Gibbs ringing. A comparison between HIPSR and MBCORR's response (before smoothing) is shown in Fig.~\ref{fig:Galactic-HI}. The significantly improved interchannel isolation of  HISPEC allows edge features to be more accurately discerned. By applying a smoothing kernel to the MBCORR data, Gibbs ringing can be suppressed; however, this decreases spectral resolution. The HISPEC modes of operation do not require a smoothing kernel to be applied.

In the presence of strong RFI, clipping may occur at the digitizer, which severely degrades the entire spectra. As HIPSR employs 8-bit sampling (256 levels), the dynamic range of the digitizer is far greater than that of the 3-level MBCORR. Combined with the improved interchannel isolation, HIPSR is capable of providing high-quality spectra in observing conditions with moderate amounts of RFI in which MBCORR cannot operate nominally.

\subsection{Extragalactic HI}

To illustrate the viability of the HIPSR system for HI surveys, this section presents an observation of a faint ($\sim$50 mJy) HI-massive galaxy, HIZOA J0836-43. This HI-massive disc galaxy lies in the Zone of Avoidance, at right ascension and declination ($\mbox{8}^{h}\mbox{36}^{m}\mbox{51}^{s}$, $-\mbox{43}^{\circ}\mbox{37}'\mbox{41}"$). It has a redshift $z=\mbox{0.036}$ and correspondingly lies at a distance of 148~Mpc. It was first detected (unresolved) in the HI Zone of Avoidance (HIZOA) survey \citep{Kraan2005}, and remains one of the most HI-massive galaxies detected to date \citep{Donley2006}. The HIZOA survey was conducted using the Parkes multibeam receiver and MBCORR, and the original archival data is still readily available\footnote{http://www.atnf.csiro.au/research/multibeam/release/}. As such, it is a fitting source to test the HIPSR system.

The source was observed for 100 s in all 13 beams. The bandpass curve was removed by using position switching between on and off source pointings of each beam. Observations of Hydra A were used to further calibrate the data and to set the flux scale. The data from all 13 beams was then combined to form an overall spectrum.

\begin{figure}
\centering
   {\includegraphics[width=0.6\columnwidth]{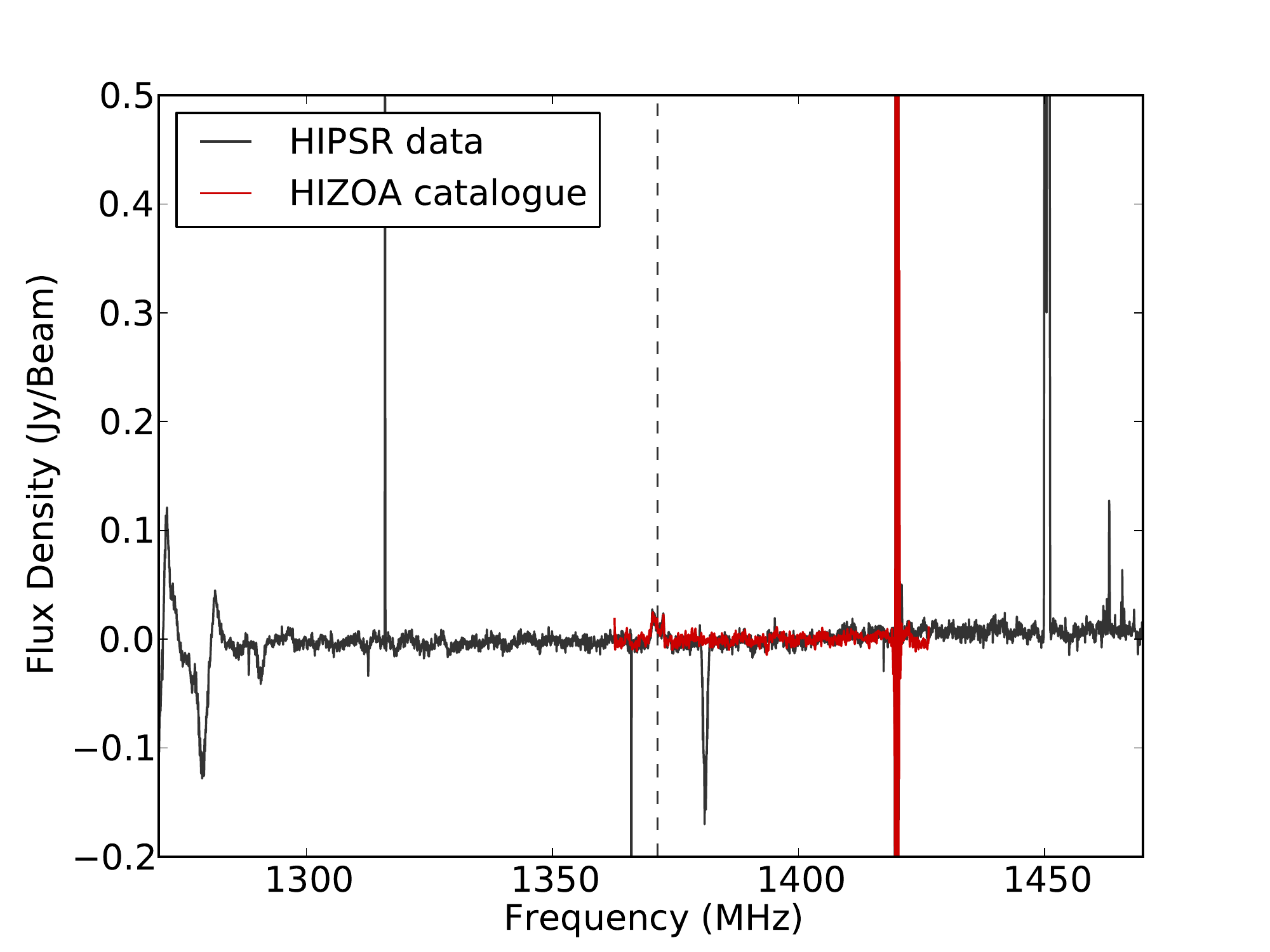}}
\caption{Calibrated HIPSR data with HIZOA archival data overlaid.\label{fig:HI-Profile-1}}
\end{figure}

\begin{figure}
\centering
    {\includegraphics[width=0.6\columnwidth]{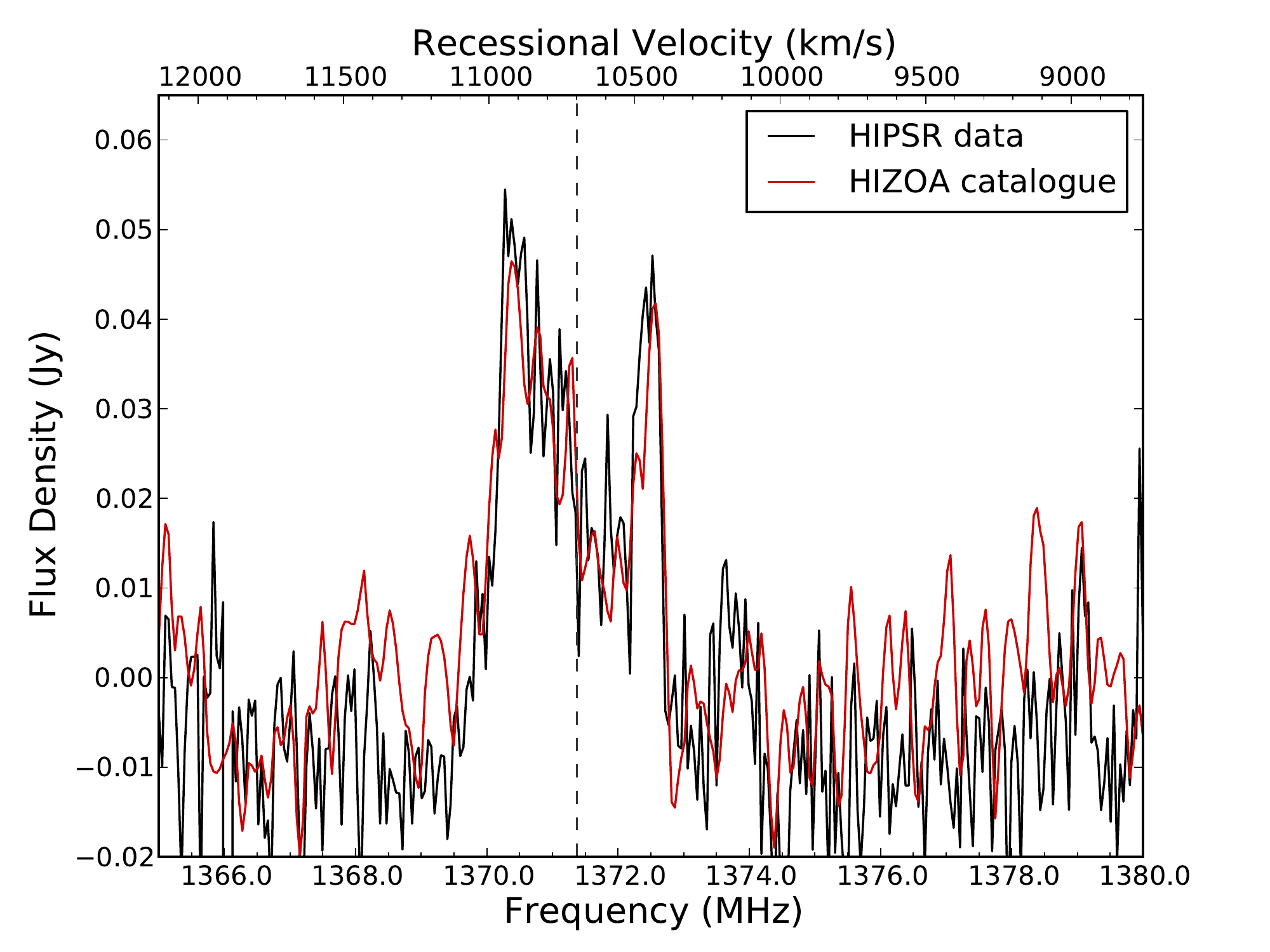}}
\caption{HIPSR first light observations of HIZOA J0836-43, showing HI profile\label{fig:HI-Profile}.}
\end{figure}

In a final calibration step, the average power level of the featureless (flat) areas of the spectrum was computed and subtracted to bring the average to 0~Jy. The final calibrated HI profile of HIZOA J0836-43 is shown in Figure~\ref{fig:HI-Profile-1} and \ref{fig:HI-Profile}, along with archival data from the HIZOA catalogue. The peak flux of the HIPSR data is 55~mJy, while the HIZOA archival data has a peak flux of 47mJy. The noise level in HIPSR spectrum is 7.9~mJy, and the HIZOA spectrum exhibits a 7.85~mJy noise level. The coefficient of determination between the two lines is $R^{2}=\mbox{0.74}$, which indicates a high degree of correlation between the observations. 

\subsection{Pulsar Observations}

As the high resolution spectrometer for both the High Time Resolution Universe (HTRU) survey and the ongoing Survey for Pulsars and Extragalactic Radio Bursts (SUPERB), the BPSR mode of HIPSR has been extensively tested in the field. BPSR data taken at Parkes after March 2012 uses the HIPSR-based system.  Here we describe some of the initial commissioning observations that were performed to assess the performance of the backend system for pulsar survey experiments.  For these tests, the channelized, detected spectra for both polarisations were de-dispersed and folded using DSPSR \citep{vb11} to form pulse profiles presented here.

\begin{figure}
\begin{centering}
\includegraphics[width=0.7\columnwidth]{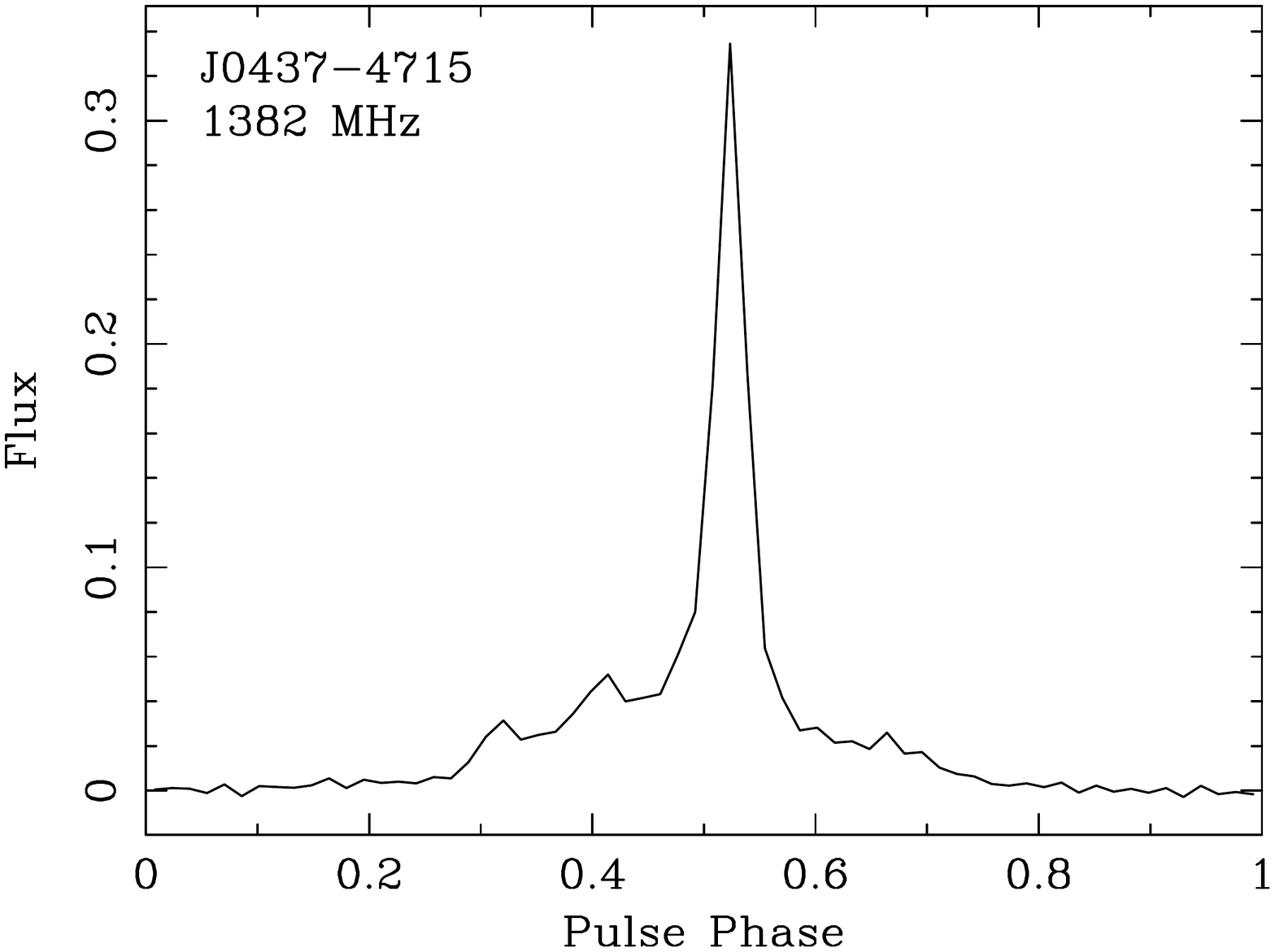}
\par\end{centering} 
\caption{Uncalibrated pulse profile of J0437-4715 
integrated for 770 seconds, observed with HIPSR during commissioning.}
\label{fig:J0437-4715_profile}
\end{figure}

The first pulsar observation performed with the HIPSR backend was of the bright millisecond binary pulsar J0437$-$4715. This pulsar was observed with the primary beam of 
the multibeam receiver for a total of 770 seconds. Fig.~\ref{fig:J0437-4715_profile} shows the pulse profile produced from this observation. BPSR was recording with a sampling interval of 64\,$\mu$s; therefore, the main peak of the pulsar, which has a full width at half maximum of around 140~$\mu$s \citep{mhb+13}, is resolved with only a few time samples.

\begin{figure}
\begin{centering}
\includegraphics[width=0.8\columnwidth]{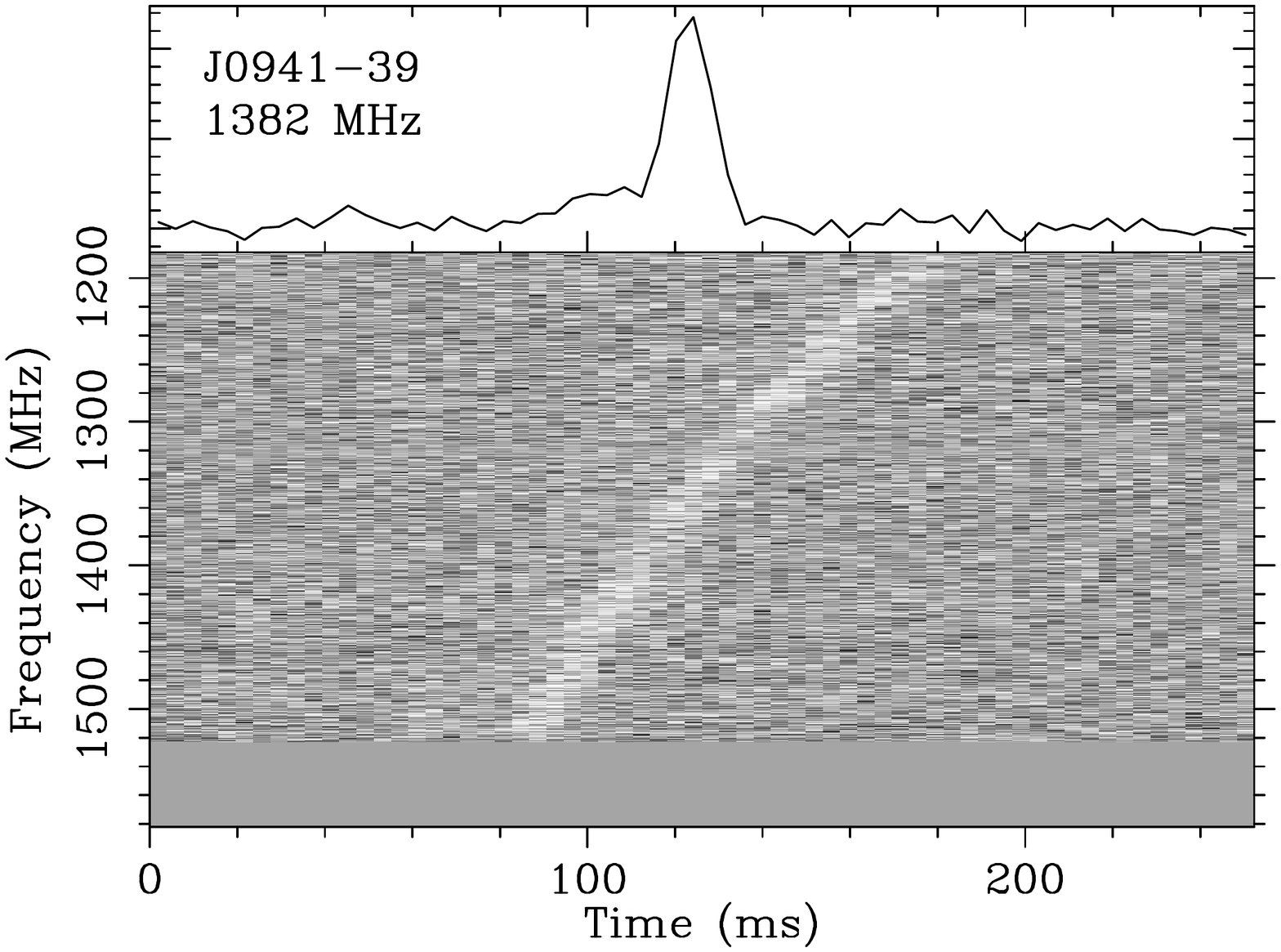}
\par\end{centering}
\caption{Single pulse from J0941$-$39.}
\label{fig:J0941-39_pulse}
\end{figure}

\begin{figure}
\begin{centering}
\includegraphics[width=0.9\columnwidth]{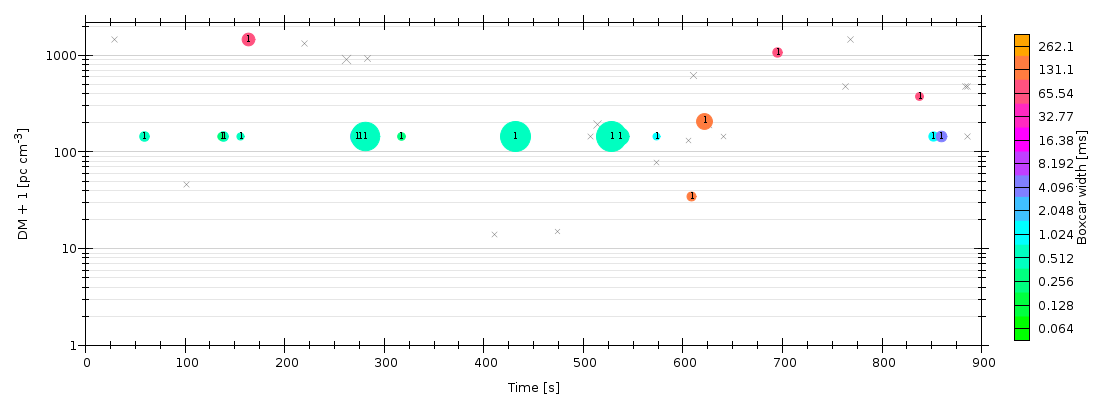}
\par\end{centering}

\caption{Detection of single pulses from J0837-24, an RRAT with DM of
142.8 in a 900 second observation using HIPSR. The size of each dot
is proportional to the Signal to Noise Ratio (SNR) of the pulse.
\label{fig:Heimdall-RRAT}}
\end{figure}

As a second test of HIPSR's high time resolution spectrometer, the intermittent pulsar J0941$-$39 was observed. In Fig.~\ref{fig:J0941-39_pulse}, a single bright pulse is shown. The bottom panel shows the dispersed pulse which has been smeared over about 100~ms. The top panel shows the pulse after the effects of dispersion have been removed. The intrinsic width of the pulse is about 4~ms.
Pulses from an intermittent pulsar, rotating radio transient \cite[RRAT;][]{mll+06} or FRB are detected in real time using {\tt Heimdall}.  Fig.~\ref{fig:Heimdall-RRAT} shows a rotating radio transient \cite[RRAT;][]{mll+06} detected in beam 01 of the multibeam receiver; plots like this are part of the real-time feedback
presented to the observer.

\section{Conclusions and future outlook}

HIPSR is a multi-purpose digital signal processing system for the Parkes 64-m telescope's 21-cm multibeam receiver; it provides a unified, multi-purpose DSP backend. HIPSR was installed at Parkes over 2011-2012, and has been in use since. As of 2016, the MBCORR backend was decommissioned, leaving HIPSR as the only digital backend for the 21-cm multibeam receiver. In the last quarter of 2016, the GP-GPU servers of HIPSR will be removed, in preparation for a future upgrade; this will mean high-time resolution modes are inoperable until a replacement GP-GPU backend is installed. The FPGA frontend and Ethernet switches will remain, and wideband 21-cm functionality will continue to be supported. 

Nevertheless, there is motivation for a full replacement of HIPSR. Since installation, the CX4 10~GbE Ethernet standard has been superceded by SFP+, and 40~GbE over QSFP+ has become common. The newer CASPER ROACH2 and SNAP FPGA boards \citep{hickish2016} use the SFP+ standard, and retains support for the iADC digitizer cards. The FPGA firmware could be ported to these platforms in a straightforward manner, so these FPGA boards are a potential avenue for future upgrades. Similarly, the GP-GPU server nodes could be replaced with current-generation equivalents. Many Ethernet switches support mixed 10 and 40~GbE, meaning fewer GP-GPU nodes could be used, or a larger fraction of bandwidth could be processed.

The most compelling motivation for an upgrade (or replacement) of HIPSR is to capture and process the entire 400~MHz bandwidth from all 26 voltage streams. Also compelling is the implementation of real-time RFI identification and excision algorithms --- for example, those described by \citet{nita2007} and \citet{kocz2010} --- decreasing the amount of data lost to interference. 

An alternative approach to increasing Parkes' L-band science capability is to install a new multi-pixel feed. Recent tests of the ASKAP Mark-II phased array feed (PAF) on Parkes \citep{chippendale2016}, have shown that PAFs are a viable technology for large single-dish telescope.  An ultra-wideband receiver, spanning 0.7-4.0~GHz is currently under development for Parkes \citep{manchester2015}, and will require a DSP backend. A unified digital backend --- similar in architecture to HIPSR --- is a potential solution for maximizing science capability at Parkes while minimizing engineering time and development costs. 

\section*{Acknowledgements}

We thanks the Parkes observatory staff for their help and support during installation and testing of the HIPSR system. This work has benefited from open source technology shared by the Collaboration for Astronomy Signal Processing and Electronics Research (CASPER). D. Price thanks J.~Hickish, G.~Foster, C.~Copley, K.~Zarb~Adami and R.~Armstrong for discussions on spectrometer implementation. HIPSR development was funded by the ARC (LE110100212).

\bibliographystyle{ws-jai}
\bibliography{journals,hipsr,psrrefs,modrefs}

\end{document}